\documentclass[fleqn,twoside]{article}
\usepackage{espcrc2}

\usepackage{graphicx}
\usepackage[figuresright]{rotating}

\newcommand{\nn}{\nonumber}
\newcommand{\wt}[1]{\widetilde{#1}}
\newcommand{\eqn}[1]{(\ref{#1})}

\newcommand{\bra}[1]{\left\langle #1\right|}
\newcommand{\ket}[1]{\left| #1\right\rangle}
\newcommand{\braket}[2]{\vev{#1 | #2}}
\newcommand{\vev}[1]{\left\langle #1 \right\rangle}

\title{
\vspace{-2cm}
\begin{flushright}
{\normalsize UTHEP-450}\\
{\normalsize UTCCP-P-115}\\
\end{flushright}
Chiral property of domain-wall fermion from eigenvalues of 4D
Wilson-Dirac Operator
\thanks{Talk presented by Y.~Taniguchi} }

\author{CP-PACS Collaboration :
  S.~Aoki\rlap,\address{Institute of Physics, University of Tsukuba, 
  Tsukuba, Ibaraki 305-8571, Japan}
  Y.~Aoki\rlap,\address{Center for Computational Physics,
    University of Tsukuba, Tsukuba, Ibaraki 305-8577,
    Japan}\thanks{present address: RIKEN BNL Research Center, 
    Brookhaven National Laboratory, Upton, NY 11973, USA}
  R.~Burkhalter\rlap,$^{\rm b}$
  S.~Ejiri\rlap,$^{\rm b}$\thanks{present address: Department of
  Physics, University of Wales Swansea, Singleton Park, Swansea SA2 8PP, UK},
  M.~Fukugita\rlap,\address{Institute for Cosmic Ray Research,
    University of Tokyo, Kashiwa 277-8582, Japan}
  S.~Hashimoto\rlap,\address{High Energy Accelerator Research Organization
    (KEK), Tsukuba, Ibaraki 305-0801, Japan}
  N.~Ishizuka\rlap,$^{\rm a,b}$
  Y.~Iwasaki\rlap,$^{\rm a,b}$
  T.~Izubuchi\rlap,\address{Institute of Theoretical Physics, Kanazawa
    University, Ishikawa 920-1192, Japan}\thanks{present address:
  Physics Department, Brookhaven National Laboratory, Upton, NY 11973, USA}
  K.~Kanaya\rlap,$^{\rm a}$
  T.~Kaneko\rlap,$^{\rm e}$
  Y.~Kuramashi\rlap,$^{\rm e}$
  V.~Lesk\rlap,$^{\rm b}$
  K.-I.~Nagai\rlap,$^{\rm b}$\thanks{present address: Theory Division, CERN, 
CH-1211 Geneva 23, Switzerland}
  M.~Okawa\rlap,$^{\rm e}$
  Y.~Taniguchi\rlap,$^{\rm a \dagger}$
  A.~Ukawa$^{\rm a,b}$ and
  T.~Yoshi\'e$^{\rm a,b}$
  }

\begin{document}

\begin{abstract}
We investigate a chiral property of the domain-wall fermion (DWF)
system using the four-dimensional hermitian Wilson-Dirac operator $H_W$.
A formula expressing the Ward-Takahashi identity quark mass $m_{5q}$
with eigenvalues of this operator is derived, which well explains
the $N_5$ dependence of $m_{5q}$ observed in previous numerical
simulations.
We further discuss the chiral property of DWF in the large volume
in terms of  the spectra of $H_W$.
\end{abstract}

\maketitle

\section{Introduction}

Recently a suitable characterization of chiral symmetry on the lattice
has appeared in the form of the Ginsberg-Wilson relation
\cite{Luscher98}. A number of numerical analysis have been carried out
for the domain wall fermion (DWF) \cite{Shamir95}, which is 
one of the simplest solutions of this relation \cite{Neuberger98,KN99}.
A chiral property of DWF is investigated numerically in two ways;
using either the chiral symmetry breaking Ward-Takahashi identity mass 
$m_{5q}$ \cite{cppacs-dwf,RBC} or the zero-eigenvalue density $\rho(0)$
of the four-dimensional hermitian Wilson-Dirac operator $H_W$
\cite{EHN99,cppacs-nagai}.  The latter is also related to the
parity-flavor breaking order parameter \cite{Aoki-phase,AKU}. 
However, conclusions are contradictory whether chiral
symmetry is realized in DWF.

In this article we derive a formula which expresses $m_{5q}$ in terms of the
eigenvalues of $H_W$, which helps to resolve the mutual inconsistency.
We further discuss the chiral property of DWF, investigating
the lattice volume dependence of $m_{5q}$ in this formula.

\section{Formula}

The anomalous quark mass $m_{5q}$ is defined through a Green function of
the chiral symmetry breaking term $J_{5q}$ \cite{Shamir95} of the
Ward-Takahashi identity:
\begin{equation}
m_{5q}=\lim_{t\to\infty}
\frac{\sum_{\vec x}\left<J_{5q}(t,{\vec x})P(0)\right>}
        {\sum_{\vec x}\left<P(t,{\vec x})P(0)\right>},
\label{eq:m5q}
\end{equation}
where $P(x)$ is the pseudo scalar density.

We try to express $m_{5q}$ in terms of the eigenvalues of $H_W$.
For this purpose we expand the Green function by inserting 
$1=\sum\ket{n}\bra{n}$ where $\ket{n}$ is the eigenstate of 
a deformed Hamiltonian $\wt{H}$ defined by 
\begin{eqnarray}
&&
\wt{H} =\log \frac{1+H'}{1-H'},
\quad
H'=H_W\frac{1}{2+\gamma_5H_W}.
\end{eqnarray}

\begin{figure}[t]
\begin{center}
\includegraphics[width=4cm]{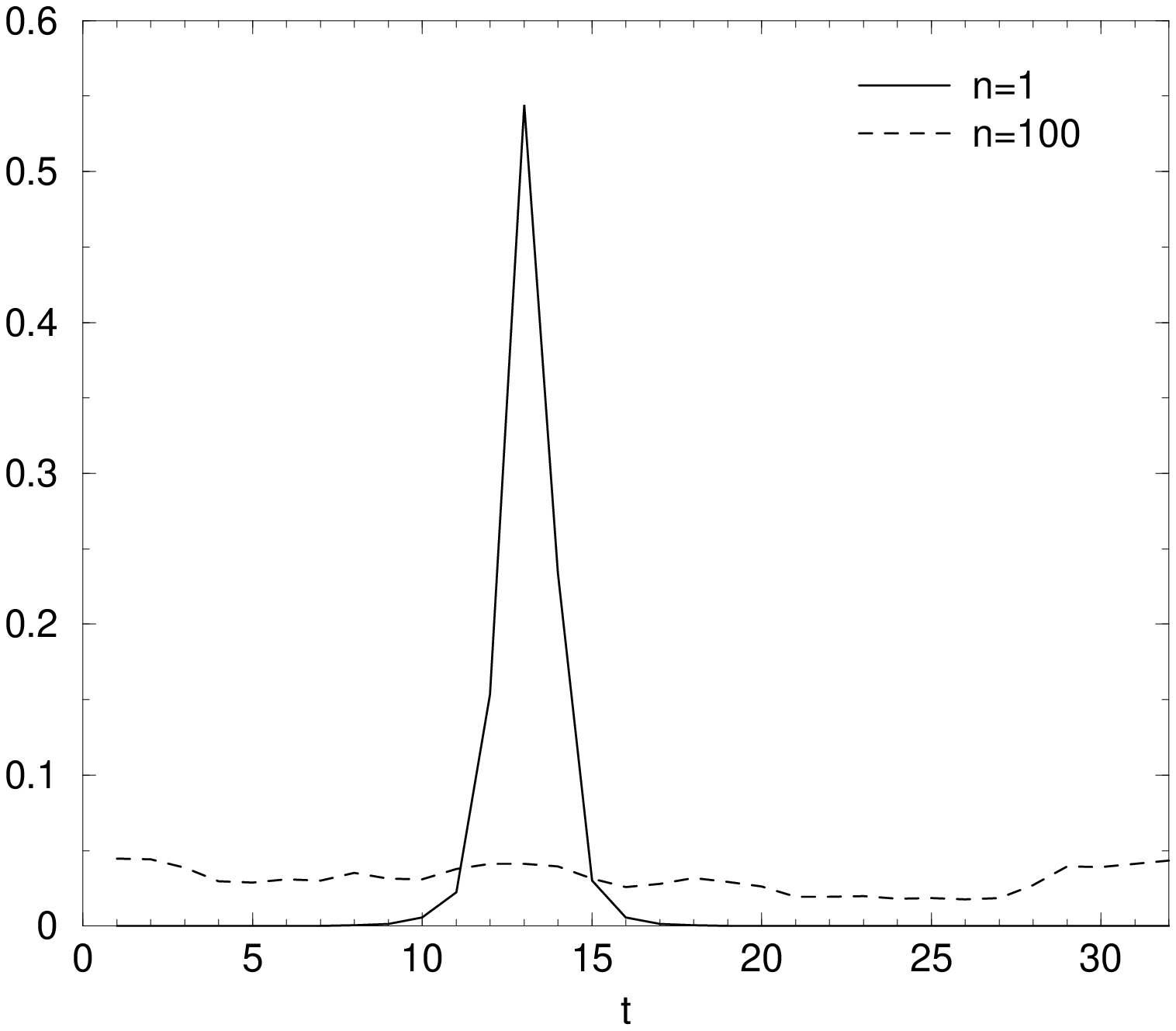}
\vspace*{-1cm}
\caption{$\sum_{\vec{x}}|\braket{x}{n}|^2$ as a function of $t$.}
\label{fig1}
\end{center}
\vspace*{-0.7cm}
\end{figure}
To simplify the resulting expression, 
we notice that the eigenvalues of $H_W$ or $\wt{H}$ are classified into 
isolated eigenvalues distributed in the lower region and 
almost continuous ones in the upper region.
As is seen in Fig.~\ref{fig1} the eigenfunctions for the isolated
eigenvalues are localized exponentially around a center and
higher excited states with continuous eigenvalues are almost plane waves, 
rapidly oscillating before absolute square is taken.

We adopt an exponential localization for the low modes and
a plane wave function for the excited modes as an
approximation.
The formula for $m_{5q}$ then becomes \cite{AT01}
\begin{eqnarray}
m_{5q}&=&
\frac{1}{12V}\Biggl(
\sum_{\rm local} \tilde h
\left(\frac{1}{2\cosh\frac{N_5}{2}\wt{\lambda}}\right)^2
\nn\\&&+
\sum_{\rm continuous}
\left(\frac{1}{2\cosh\frac{N_5}{2}\wt{\lambda}}\right)^2
\Biggr),
\label{formula}
\end{eqnarray}
with $V$ the space-time volume, 
$\wt{\lambda}$ an eigenvalue of $\wt{H}$ and
$\tilde h$ being related to
the exponential width $\delta$ of the localized eigenfunctions as
$\tilde h\!=\!(4\delta)^4$.

\section{Numerical results}

\begin{figure}[t]
\begin{center}
\includegraphics[width=5cm]{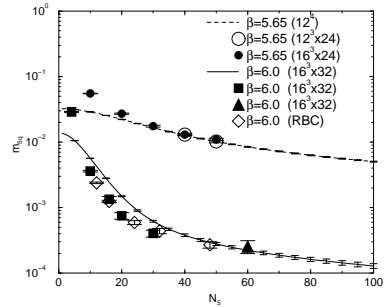}
\vspace*{-1cm}
\caption{$m_{5q}$ as function of $N_5$ for the plaquette action.
Lines are our formula. RBC is a data from \cite{RBC}.}
\label{fig2}
\end{center}
\vspace*{-0.3cm}
\end{figure}
\begin{figure}[t]
\begin{center}
\vspace*{-0.7cm}
\includegraphics[width=5cm]{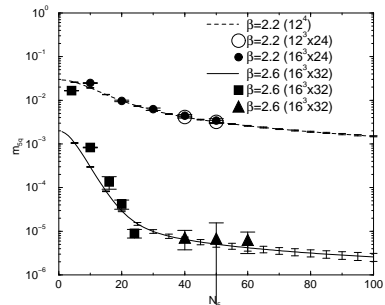}
\vspace*{-1cm}
\caption{$m_{5q}$ as function of $N_5$ for the RG action.}
\label{fig3}
\end{center}
\vspace*{-0.8cm}
\end{figure}

In Figs.~\ref{fig2} and \ref{fig3} we compare $m_{5q}$ 
obtained with \eqn{formula} (lines) with those directly
calculated in simulations, both for the plaquette and the RG improved 
gauge actions, where circles and squares are from \cite{cppacs-dwf,RBC} 
and triangles are from this work.

\begin{figure}[t]
\begin{center}
\includegraphics[width=5cm]{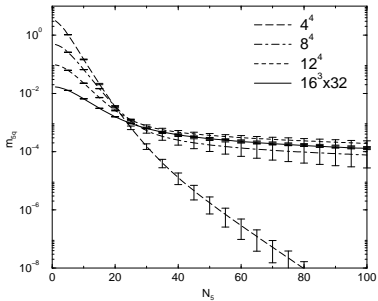}
\vspace*{-1cm}
\caption{Volume dependence of $m_{5q}$ for the plaquette action at
 $\beta=6.0$.} 
\label{fig4}
\end{center}
\vspace*{-1.0cm}
\end{figure}

The right hand side of (\ref{formula}) is calculated as follows. 
We evaluate eigenvalues of the Wilson Dirac operator $H_W$ using the
Lanczos method. 
The eigenvalue $\wt{\lambda}$ of the deformed Hamiltonian is evaluated
from that of $H_W$ perturbatively for the isolated eigenvalues and
directly for the continuous ones assuming the free theory relation.
The separation between the isolated and the continuous eigenvalues is made 
by inspection of the eigenvalue distribution, and 
$\wt{h}$ is treated as a free parameter to fit the data.
For example, at $\beta\!=\!2.6$ of the RG action, 
we adopt the number of localized eigenvectors $n_l\!=\!20$, 
and $\wt{h}\!=\!200$. 
Summing up contributions from each eigenvalue we obtain $m_{5q}$ as a
function of $N_5$.
Errors are calculated by a single elimination jackknife procedure for
$100$ configurations. 

As is seen in these figures our formula well explains the $N_5$ dependence
of the anomalous quark mass both in strong and weak coupling regions 
for the two gauge actions. 
An important problem in the previous numerical investigations is
whether $m_{5q}$ contains a constant term in the large $N_5$ limit.
Now we can conclude that the exponential dumping still holds even at
large $N_5$ but with a small decay rate, which reflects the magnitude of 
a few small eigenvalues.
This is a reasonable behavior for finite lattice volume where 
exact zero eigenvalues are absent.

A bending behavior of $m_{5q}$ for weak coupling around $N_5\!\sim\!20$ 
can be explained with our formula as follows.
For small $N_5$ the continuous modes, which dominate in number,
mainly contribute in \eqn{formula}, giving a steep exponential decay.
For large $N_5$, on the other hand, only a few of the small isolated 
eigenvalues contribute, so that the dumping rate of $m_{5q}$ becomes small.

\section{Volume dependence of $m_{5q}$}

Although an exponential decay of $m_{5q}$ in $N_5$ seems to
hold in DWF at finite volume,
there remains a crucial question whether this remains so, and hence 
chiral symmetry is exactly realized with DWF, in the infinite volume limit.
Indeed $m_{5q}$ decreases with some power of $1/N_5$ if the number of 
small eigenvalues increases linearly with the volume $V$
\cite{AT01}, in which case chiral symmetry is not realized.

\begin{figure}[t]
\begin{center}
\includegraphics[width=5cm]{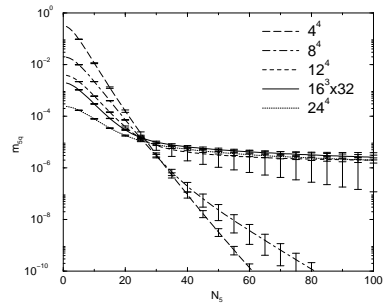}
\vspace*{-1cm}
\caption{Volume dependence of $m_{5q}$ for the RG action at
 $\beta=2.6$.} 
\label{fig5}
\end{center}
\vspace*{-0.9cm}
\end{figure}

To examine this question we plot the volume dependence of $m_{5q}$
at a weak coupling in Figs.~\ref{fig4} and \ref{fig5},
using the formula \eqn{formula} with $n_l$ and $\tilde h$ obtained
in the previous section.
Almost no volume dependence is observed for $V\ge12^4$ at large $N_5$ 
for both actions, and a similar behavior holds also
in the strong coupling region.

It is difficult, however, to distinguish a slow exponential decay 
seen in Figs.~\ref{fig4} and \ref{fig5} from a power of $1/N_5$
at $N_5\sim100$.
A distinction becomes possible only for $N_5\gg100$,
which effectively corresponds to $N_5\to\infty$
for the current lattice volumes, $V\simeq 12^4$ -- $24^4$, so
data at much larger volumes are needed to establish
the stability of an exponential decay against the increase of volume.
Therefore the current data do not allow definite conclusions on the
chiral property of DWF in the infinite volume limit.  
Further investigations  will be required to answer the question.

\vspace{2mm}
This work is supported in part by Grants-in-Aid of the Ministry of Education 
(Nos.~10640246, 
10640248, 
11640250, 
11640294, 
12014202, 
12304011, 
12640253, 
12740133, 
13640260  
).

\newcommand{\J}[4]{{#1} {#2}, #3 (#4)}
\newcommand{\AP}{Ann.~Phys.}
\newcommand{\CMP}{Commun.~Math.~Phys.}
\newcommand{\IJMP}{Int.~J.~Mod.~Phys.}
\newcommand{\MPL}{Mod.~Phys.~Lett.}
\newcommand{\NP}{Nucl.~Phys.}
\newcommand{\NPSup}{Nucl.~Phys.~B (Proc.~Suppl.)}
\newcommand{\PL}{Phys.~Lett.}
\newcommand{\PR}{Phys.~Rev.}
\newcommand{\PRL}{Phys.~Rev.~Lett.}
\newcommand{\PTP}{Prog. Theor. Phys.}
\newcommand{\Suppl}{Prog. Theor. Phys. Suppl.}
\newcommand{\RMP}{Rev. Mod. Phys.}

\end{document}